\documentclass[twocolumn]{aastex631_modif}

\accepted{\today}

\shorttitle{School students observe Venus with IRTF}
\shortauthors{Peralta et al.}

\begin{document}

\title{Secondary School Students observe Venus with NASA Infrared Telescope Facility (IRTF)}

\correspondingauthor{Javier Peralta}
\email{jperalta1@us.es}

\author[0000-0002-6823-1695]{Javier Peralta}
\affiliation{Facultad de Física, Universidad de Sevilla \\
Avenida de la Reina Mercedes s/n \\
Sevilla, Spain}

\author[0000-0001-9607-0190]{Juan A. Prieto}
\affiliation{Colegio Huerta de la Cruz \\
Calle Vicente de Paul, 7 \\
Algeciras (C{\' a}diz), Spain}

\author{Pilar Orozco-S{\' a}enz}
\affiliation{Colegio Huerta de la Cruz \\
Calle Vicente de Paul, 7 \\
Algeciras (C{\' a}diz), Spain}

\author{Jes{\' u}s Gonz{\' a}lez}
\affiliation{Colegio Huerta de la Cruz \\
Calle Vicente de Paul, 7 \\
Algeciras (C{\' a}diz), Spain}

\author{Gonzalo Trujillo}
\affiliation{Colegio Huerta de la Cruz \\
Calle Vicente de Paul, 7 \\
Algeciras (C{\' a}diz), Spain}

\author{Luc{\' i}a Torres}
\affiliation{Colegio Huerta de la Cruz \\
Calle Vicente de Paul, 7 \\
Algeciras (C{\' a}diz), Spain}

\author{Alberto S{\' a}nchez}
\affiliation{Colegio Huerta de la Cruz \\
Calle Vicente de Paul, 7 \\
Algeciras (C{\' a}diz), Spain}

\author{Manuel Arnedo}
\affiliation{Colegio Huerta de la Cruz \\
Calle Vicente de Paul, 7 \\
Algeciras (C{\' a}diz), Spain}




\begin{abstract}
Astronomy and astrophysics are regarded as highly motivating topics for students in primary and secondary schools, and they have been a recurrent and effective resource to inspire passion about science. In fact, during the last years we have witnessed a boost of facilities providing small robotic telescopes for teachers and students to remotely undertake their own observing projects. A step forward is presented here, where we describe the experience of secondary school students attending professional observations of Venus at NASA's Infrared Telescope Facility (IRTF) and, in a second observing run, conducting the observations by themselves. In addition to quickly mastering the basic operation of the control software for the SpeX instrument, the students successfully performed different types of data acquisition, including drift scan imaging.
\end{abstract}


\section{Introduction} \label{sec:intro}
Astronomy is one of the oldest sciences known. For millennia, it has captivated most of the cultures in the world, and it yet remains at the forefront of the attention and interest of public \citep{Bailey2003}. Many teachers have used astronomy to counter the problem that many students find the science content of middle years of schooling uninteresting \citep{Salimpour2021}. In light of the global push to get students engaged in science and technology, many aspects of astronomy have become popular and introduced in school curricula for decades \citep{Lelliott2010}, leveraging the many examples in astrophysics with direct links to Physics, Chemistry, Mathematics and even Biology \citep{Salimpour2021}.\\

\begin{figure*}[ht!]
\centering
\includegraphics[width=0.9\textwidth]{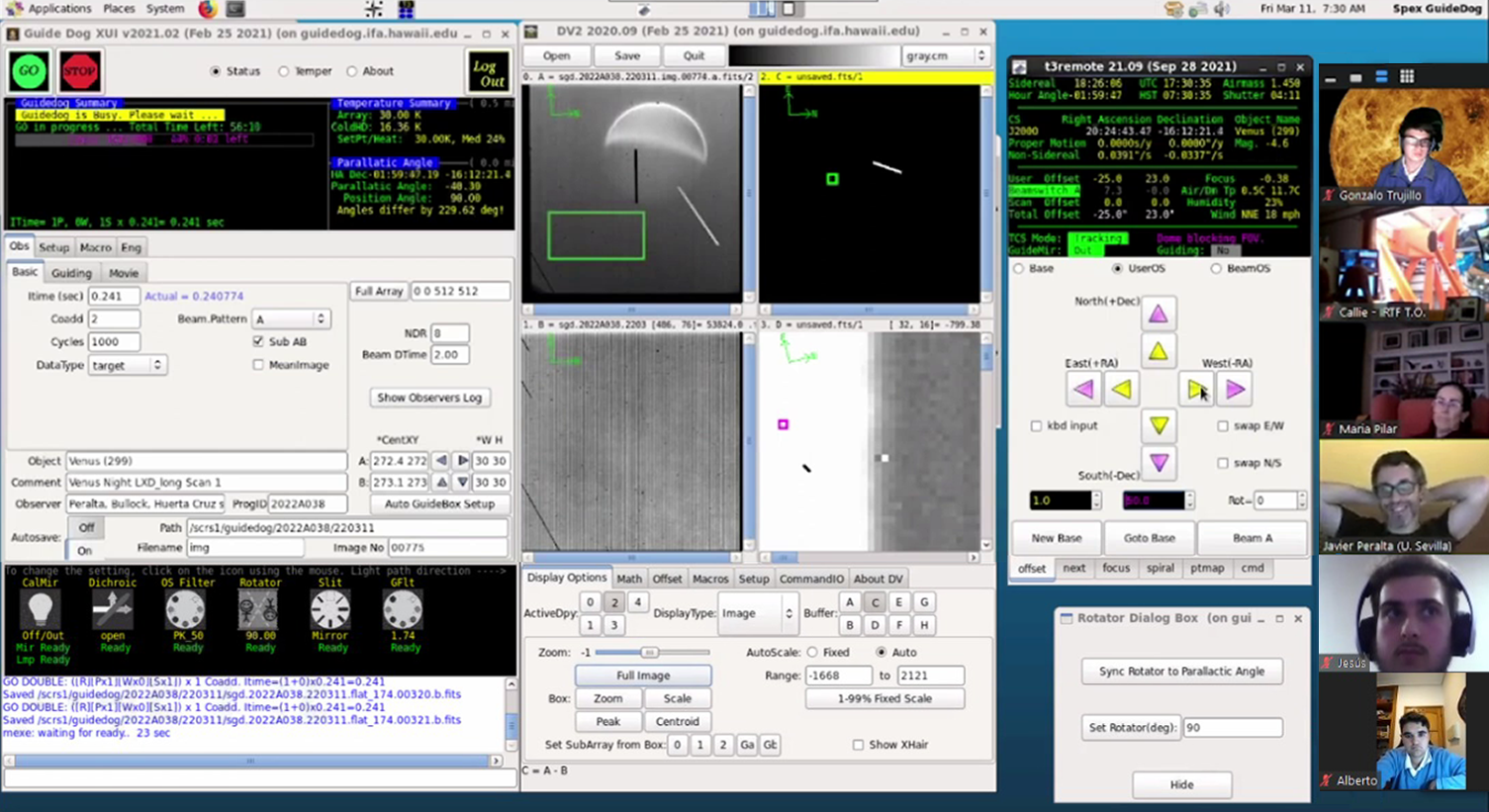}
\caption{Capture of screen during observing run conducted by the students in 13 of March 2022.
\label{fig:observation}}
\end{figure*}

In this context, there has been an significant growth in the number robotic telescopes with observing time fully or partially devoted to educational purposes and friendly user interface to ease the remote control by school students \citep{Gomez2017}. This has allowed school teachers to easily implement inquiry-based learning approaches and students to have authentic science experiences, start international collaborations and even make discoveries and publish the results \citep{Salimpour2018,Fitzgerald2018}.\\

In this work, we describe how a team of secondary school students performed professional observations of the planet Venus with the spectrograph and imager SpeX \citep{Rayner2003} at the National Aeronautics and Space Administration Infrared Telescope Facility (NASA/IRTF) on Mauka Kea (Hawaii). Venus is a captivating target since it exhibits the consequences of a runaway greenhouse effect and it has been recently in the spotlight of the search for life \citep{Greaves2021}.\\

\section{Methodology} \label{sec:method}
The three participating students (J. Gonz{\' a}lez, A. S{\' a}nchez and G.~Trujillo) were aged 15-16 and studied at \textit{Huerta de la Cruz}, a catholic school located in the city of Algeciras (Spain). These students volunteered to carry out a project on Venus \citep[see \textit{work}]{Peralta2023zenodo}, motivated and coordinated by J.~A. Prieto and P. Orozco-S{\' a}enz, professors with successful experience engaging students with scientific research \citep{Prieto2020,Prieto2022sons,Orozco2021}. The objectives of this work comprised acquiring a basic knowledge of Venus and its atmosphere, a discussion about its habitability, and knowing the methods for observing Venus \citep[see \textit{log file}]{Peralta2023zenodo}.\\ 

With regards to the latter objective, the students were invited to participate in professional observations at NASA/IRTF between January and March of 2022. Two programs for observing Venus were scheduled\footnote{\url{http://irtfweb.ifa.hawaii.edu/observing/schedule.php}}: programs 2022A058 (with E.~F. Young as Principal Investigator) and 2022A038 (J. Peralta as Principal Investigator). Although with different scientific goals, both programs shared similar techniques of Venus data acquisition with SpeX \citep[see \textit{work}]{Peralta2023zenodo}: sets of Venus images with several filters using the guide camera (hereafter \textit{guidedog}), hyperspectral data with the slit of the spectrograph (hereafter \textit{bigdog}) scanning from north to south the disk of Venus, guiding on a star to obtain calibration spectra, or inference of flats, darks and bias. The full imagery dataset from these two programs has already been employed in recent Venus research \citep{Peralta2023AyA}, and it will become publicly available at NASA/IPAC Infrared Science Archive.\\

Prior to undertaking observations of Venus with SpeX, the students used the bibliography to became introduced to the clouds of Venus \citep{Peralta2019Icarus}, the SpeX instrument \citep{Rayner2003}, and the graphical user interfaces to control \textit{guidedog} (\textit{Guidedog X-windows User Interface} or GXUI, and \textit{Guidedog Data Viewer} or GDV) and the T3 remote TCS widget to adjust the pointing coordinates and the focuser \citep{Rayner2021}. After this literature review, the students benefited from a training session by remotely attending an observing run during 17 of February 2022 (2022A058). Finally, the students joined a last run during 11 of March 2022 (2022A038) to conduct the Venus observations by themselves (see Figure \ref{fig:observation}).\\

\newpage

\section{Experience}\label{sec:result}
During the observing run of 11 of March 2022, the students conducted the remote observations of Venus with IRTF/SpeX for about 2 hours, and 1 hour was recorded\footnote{Video available upon reasonable request and for educational purposes only.}. Jes{\' u}s Gonz{\' a}lez was selected to make the VNC connection and control GXUI, GDV and the T3 remote TCS widget, supervised by Javier Peralta. In addition to quickly understand most of the explications, J. Gonz{\' a}lez also performed the following operations with success: 
\begin{itemize}
    \item \textit{Fast acquisition of images with different filters and integration times.}
    \item \textit{Acquisition of hyperspectral data.} The student used T3 remote to introduce corrections for the position of the slit during the drift scan of the night and day side of Venus.
    \item \textit{Acquisition of images/spectra of the sky.} The student used T3 remote to shift the slit to a location of the sky at a convenient angular separation.
    \item \textit{IR guiding with SpeX guider/slit viewer.} The student properly employed GXUI and GDV to move a target star into the telescope A beam box and start the guiding with the Auto GuideBox Setup \citep{Rayner2021}.
\end{itemize}

\section{Conclusions}
We have shown in this work that a group of motivated secondary school students have been able to understand basic and more elaborated strategies to observe a planet of the solar system, and also learn to use a complex telescope interface to undertake the observations scheduled for a professional program. Encouraged by the positive experience, we plan to expand this recognised work \citep{Prieto2022cea} with new research projects adapted to secondary education which may include an active participation in the design of professional observing proposals and/or leading novel research under the supervision of postdoctoral researchers, an initiative successfully implemented in other projects \citep{Dunn2018}.\\

\section*{Author Contributions}
Conceptualization, software, resources, data curation, writing—original draft preparation and funding acquisition, by J.P.; methodology, writing—review and editing, supervision and validation, by J.P., J.A.P. and P.O-S.; investigation, visualization and formal analysis, by J.G., M.A., A.S. and G.T.\\

\begin{acknowledgments}
 J.P. thanks funding by the program EMERGIA, Junta de Andalucía (Spain), Grant EMERGIA20{\_}00414. J.A.P. and P.O-S. acknowledge the support from Asociaci{\' o}n Amigos de la Ciencia, Diverciencia. We also thank \textit{E.~F. Young} and \textit{M.~A. Bullock} for allowing the students to attend one of their observing runs. We acknowledge the support from \textit{Callie Matulonis} (Telescope Operator during the run with students) and the rest of staff at the Infrared Telescope Facility, which is operated by the University of Hawaii under contract 80HQTR19D0030 with the National Aeronautics and Space Administration. 
\end{acknowledgments}


\vspace{5mm}
\facilities{IRTF(SpeX)}



\bibliographystyle{aasjournal}



\end{document}